# Reducing water entry impact forces

Nathan B. Speirs[1], Jesse Belden[2], Zhao Pan[1], Sean Holekamp[2], George Badlissi[3], Matthew Jones[1], and Tadd T. Truscott[1]†

[1]Department of Mechanical and Aerospace Engineering, Utah State University, Logan, UT 84322, USA

[2]Naval Undersea Warfare Center Division Newport, 1176 Howell Street, Newport, Rhode Island 02841, USA.

[3]Department of Ocean Engineering, University of Rhode Island, Narragansett, Rhode Island 02882, USA.

The forces on an object impacting the water are extreme in the early moments of water entry and can cause structural damage to biological and man-made bodies alike. These early-time forces arise primarily from added mass, peaking when the submergence is much less than one body length. We experimentally investigate a means of reducing impact forces on rigid spheres by making a jet of water strike the quiescent water surface prior to the object impacting. The water jet accelerates the pool liquid and forms a cavity into which a sphere falls. Through on-board accelerometer measurements and high speed imaging, we quantify the force reduction compared to the case of a sphere entering a quiescent pool. Finally, we find the emergence of a critical jet volume required to maximize force reduction; the critical volume is rationalized using scaling arguments informed by near-surface particle image velocimetry (PIV) data.

## 1. Introduction

Free surface impact has been investigated for over a century (Worthington & Cole 1900) with most studies examining solid or liquid impact on a quiescent pool. In this study, we examine the phenomenon of a solid body descending through a transient air cavity that is formed by a liquid jet, as shown in Fig. 1. The impact of the liquid jet greatly alters the flow field into which the sphere enters, which in turn dramatically changes the forces on the sphere during entry. In this paper, we examine these forces and find that the very initial impact force can be greatly reduced when the sphere impacts the bottom of a jet cavity.

Prior research on solid impact on a free surface informs our study on the impact forces. One of the first to study the forces during free surface impact was Thompson (1928), who experimentally investigated the maximum pressure on sea plane floats during landing. Von Karman (1929) followed up this study by developing a formula to apply Thompson's experimental results to different shaped floats and impact velocities and was one of the first to model the impact force using added or virtual mass. Shiffman & Spencer (1945) studied the impact of spheres on water up to a submergence depth of one radius and mathematically predicted the drag coefficient as a function of submergence depth using added mass arguments. They found that the maximum drag coefficient occurs when the sphere is submerged between ten and twenty percent of its radius. Others have shown similar trends for other geometries, with added mass being the dominant source of large peak forces for small body submergence (May 1975; Grady 1979). Further theoretical developments on modeling the very initial impact force have been reviewed by Korobkin & Pukhnachov (1988) with Miloh (1991) and Faltinsen & Zhao (1997) making significant contributions since. Moghisi & Squire (1981) experimentally validated the work of Shiffman & Spencer (1945) for low viscosity liquids and impact velocities between 1 and 3 m/s. They also found that the impact force varies with the square root of the depth for depths less than ten percent of the radius. Further work on the initial impact force was performed by Bodily *et al.* (2014) who studied the water entry of slender axisymmetric bodies. They showed that the impact force is a function of nose geometry.

† Email address for correspondence: taddtruscott@gmail.com

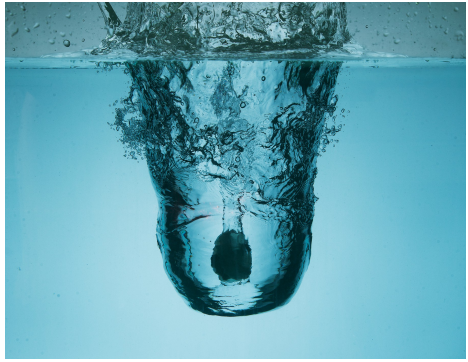

FIGURE 1. A 25 mm radius sphere impacts at the bottom of an air cavity previously formed by the impact of jet of water on a quiescent pool.

The forces experienced after the very initial stages of impact depend on whether the sphere pulls air under the surface with it or enters without air entrainment (Truscott *et al.* 2012). In cases where cavities form as the falling sphere comes in contact with the water, the water is repelled away from the sphere near the sphere's equator and an air cavity forms in its wake. If the sphere enters the water without entraining air the water will travel up the sides of the sphere meeting at the sphere's apex, thus preventing entrainment. While cavity formation can be suppressed at low velocity if the static contact angle is less than 90° (Duez *et al.* 2007) and the sphere is smooth (Zhao *et al.* 2014), cavities always form with sufficiently large sphere impact velocity ($U_o \gtrsim 7.3$ m/s in water) which decreases as the contact angle or roughness increase . Once a cavity forms, the balance between the inertial, gravitational, and surface tension forces, described by the Bond, Weber, and Froude numbers (defined below), dictate its dynamics and cause it to take on one of four shapes or regimes (Aristoff & Bush 2009). The two applicable cavity regimes for this study are the deep seal regime, in which hydrostatic pressure forces the cavity to close approximately halfway between the sphere and water surface, and the surface seal regime in which the splash collapses inwards sealing off further air flow into the cavity (Aristoff & Bush 2009). In the current study, cavities always form when impacting a quiescent pool due to the high surface roughness of the sphere and both deep and surface seals are seen.

Other studies have focused on the forces experienced after the initial impact. May & Woodhull (1948) found the average drag coefficient of steel spheres during the entrance cavity phase, while Shepard *et al.* (2014) studied the effect of sphere density on the drag coefficient for the same time phase. Truscott *et al.* (2012) showed that the forces during these later stages of impact are very unsteady and depend on whether the sphere forms a cavity. In non-cavity forming cases vortices shed in the wake of the sphere cause large impulses in the sphere acceleration. When cavities form the trailing air bubble suppresses vortex shedding and the sphere experiences other forces caused by the pinch-off event (Bodily *et al.* 2014). Other studies on water impact include: Glasheen & McMahon (1996) who studied the impact forces of circular disks to understand how basilisk lizards and shore birds run along the water surface, Baldwin (1971) who studied cones, and Tveitnes *et al.* (2008) who studied wedges.

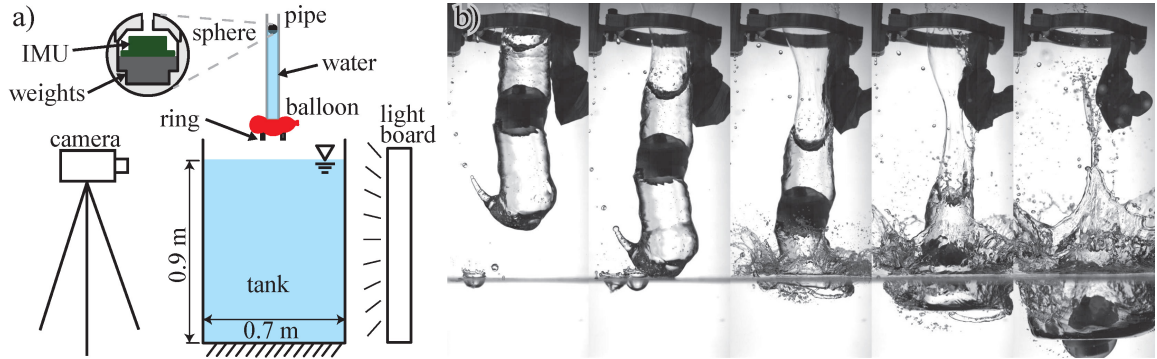

FIGURE 2. A schematic of the experimental setup is shown in a) with a zoomed in view of the IMU nestled in the interior. b) When the balloon pops the jet and sphere fall towards the pool and the jet forms a cavity into which the sphere falls (time between images is 20 ms). In the upper part of the images the ring and popped balloon can be seen.

In this study we focus on a method for reducing the initial, impulsive impact force experienced by a sphere during water impact. As described above, the body of literature shows that this early time force, occurring before the sphere is fully submerged, is predominately caused by the sphere having to accelerate the surrounding water; i.e., added mass. Here we suggest that the impact force can be reduced by accelerating a volume of water just below the free-surface prior to sphere impact. To test this experimentally, we allow a jet of water to strike the surface prior to the sphere impacting. On-board measurements of acceleration using custom inertial measurement units (IMUs) confirm that the large initial impact force is reduced, and that the subsequent forces and cavity dynamics throughout entry are also altered. We investigate this effect over a range of impact velocities and water jet lengths, and place our findings in the context of prior work on rigid sphere impact on a quiescent pool.

## 2. Experimental setup and description

Figure 2a shows the setup used for this study. A polycarbonate pipe of inner diameter 51 mm is held above a tank of water and the bottom opening of the pipe is sealed by smashing an inflated party balloon against it using a metal ring. The pipe is filled with water to the desired height and then a 50 mm diameter sphere is placed at the top of the water in the pipe. When the balloon is popped it quickly moves out of the way and the water and sphere fall towards the pool surface as seen in Fig. 2b. As the jet impacts the pool it spreads on the surface and forms an air cavity into which the sphere falls. The sphere impacts the bottom of the cavity and the impact event is viewed at 1000 fps with a high-speed camera imaging below the pool surface with diffuse back lighting. In some cases a second synchronized camera views the jet and sphere above the pool surface. The same sphere is also dropped without a jet and impacts a quiescent pool. Measurements are taken from these videos to find the sphere impact velocity, cavity velocity, and the length of jet in front of the sphere.

The sphere consists of an outer shell, weights, and an inertial measurement unit (IMU), as shown in Fig. 2a. The outer shell of the sphere is 3D printed in two parts using Vero plastic. This provides a hydrophilic surface with wetting angle $\theta = 80 \pm 8°$ and surface roughness $R_z = 7.2 \pm 1.2$ $\mu$m at 95% confidence. The steel weights are placed in the lower half of the sphere with the IMU firmly attached to them. This helps the sphere to fall inline with its vertical axis and to minimize the sphere rotation during free fall and impact; the magnitude of the total acceleration vector is computed from measurements in the three axes and is reported herein. The two pieces of the sphere are pressed together and the seam and top hole sealed with Colorimetrics gray putty tape to prevent water from entering. The seam between the two pieces of the sphere shell is located about two thirds of the sphere radius from the bottom of the sphere so as to minimize its influence on the dynamics of the water impact event. The specific gravity of the sphere as a whole is 2.253 ±0.007.

The IMU was built in house and has two three-axis-accelerometers, that separately record each impact event at 1000 Hz. The low range accelerometer is set to a maximum range of ±16 g and the high range accelerometer is set to a maximum range of ±100 g. When possible the data from the low range accelerometer is used as it results in less noise, but the measurements of the two separate accelerometers are comparable. Because the sphere experiences small rotations during free fall and impact the three components of acceleration are summed and the magnitude of the acceleration vector is reported herein.

Using the setup described the impact velocity at the cavity bottom or quiescent pool surface, $U_o$, was changed from 1.83 to 9.34 m/s by varying the drop height. The length of the jet impacting in front of the sphere, $L_j$, varied from 0 to 55 cm. This resulted in nondimensional parameters with the following ranges: $Re = \rho U_o R_s/\mu$ between 40,000 and 200,000, $We = \rho U_o^2 R_s/\sigma$ between 1,100 and 31,000, and $Fr = U_o/\sqrt{gR_s}$ between 3.6 and 18.9, where $\rho$ is the liquid density, $R_s = 25$ mm is the radius of the sphere, $\mu$ is the dynamic viscosity of the liquid, $\sigma$ is the liquid-air surface tension, and $g$ is the acceleration of gravity. Some limitations of the setup are that the sphere and jet radii must be equal and that the drop height of the cases with jets could not be increased above 4 m as the jet front becomes more distorted with increased falling distance.

### 2.1. Uncertainty

Uncertainty in all measurements is calculated and the uncertainty bands in the figures represent the 95% confidence interval of the measurement (Coleman & Steele 2009). The uncertainty in calculated variables was often found to scale monotonically with the variable. Where applicable, two or more bands are placed on the extremes of the figure axes or with the data set to show how the uncertainty scales (e.g., Fig. 4b), when only one band is present the mean uncertainty is shown.

## 3. Results and discussion

Figure 3a shows an image sequence of a sphere impacting a quiescent pool of water at $U = 4.39$ m/s with the corresponding acceleration of the sphere shown in Fig. 3c. In the very early stages of impact the sphere accelerates a portion of the surrounding water (added mass) (Shiffman & Spencer 1945), which causes a large, but short lived peak in the acceleration of the sphere (Fig. 3c, $t = 0$ to 0.01 s). A cavity then forms expanding downwards into the pool. At 25 ms after impact the splash crown domes over with no immediately noticeable influence on the sphere acceleration. At approximately 100 ms a deep seal occurs causing ripples and volume oscillations in the lower portion of the cavity which give rise to the oscillations seen in the sphere acceleration with approximate amplitude of 0.45 g (Grumstrup et al. 2007; Bodily et al. 2014) At approximately 175 ms a bubble sheds from the lower portion of the cavity increasing the amplitude of the oscillations in the sphere acceleration to about 0.59 g.

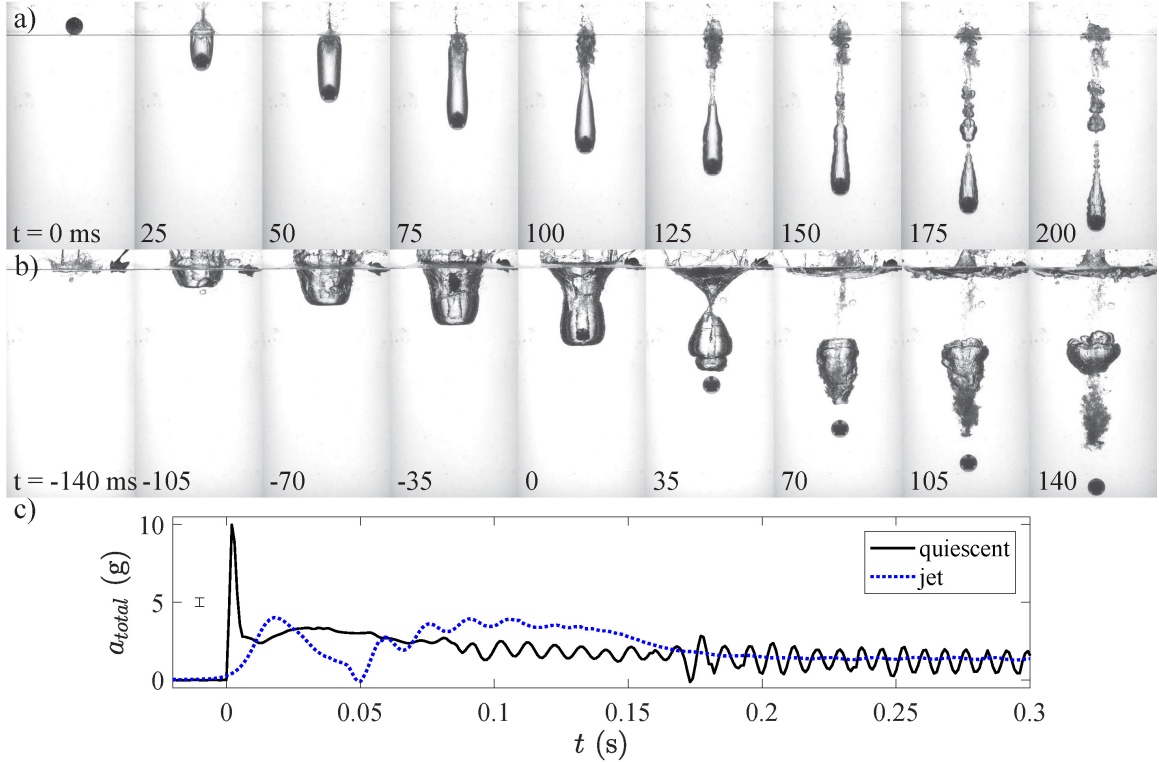

FIGURE 3. a) A 50 mm diameter sphere impacts a quiescent pool surface with velocity $U = 4.39$ m/s forming a subsurface air cavity, that experiences surface seal, deep seal and cavity shedding at around, 25, 100, and 175 ms respectively. b) A 50 mm diameter water jet impacts a pool surface forming a subsurface air cavity. At $t = 0$, the 50 mm diameter sphere impacts the bottom of the jet cavity at velocity $U = 4.35$ m/s without forming a cavity. c) The total acceleration $a_{total}$ vs. time $t$ is plotted for the sphere impacting a quiescent surface in a) and for the sphere impacting behind a jet in b). See supplemental movies 1 & 2.

If a sphere is placed inside a falling jet of water, the jet impacts the pool prior to sphere impact and forms an air cavity into which the sphere falls. This is shown in Fig. 3b for approximately the same impact velocity as for the quiescent impact case shown in a. Immediately after impact the acceleration of the sphere increases but not to as large a value as seen in the quiescent impact case (Fig. 3c, $t = 0$ to 0.05 s). The sphere enters the pool without forming a cavity because the sphere is already immersed inside the jet. The cavity previously formed by the water jet collapses in a deep seal at 35 ms after sphere impact. The large bubble formed by the pinch-off event oscillates leading to oscillations in the sphere acceleration, which for the case shown begins with an amplitude of about 1 g and decrease exponentially as the sphere descends away from the bubble. These oscillations are superimposed on the increase of acceleration from t = 0.05 to 0.15, which is caused by a vortex shed from the sphere as discussed by Truscott et al. (2012).

As the jet significantly reduces the maximum impact force experienced by the sphere during the very initial stage of impact (Fig. 3c $t = 0 - 0.01$ s and $t = 0 - 0.05$ s) we now focus on this time period and, in particular, the maximum acceleration during this early stage. The maximum measured acceleration $a_{max}$ is normalized by $g$ and plotted as a function of $Fr$ in Fig. 4a. For the sphere impacting on a quiescent pool, the maximum acceleration increases quadratically with $Fr$. We can predict this behavior from a force balance including total drag, $a_{max} = \frac{1}{2} C_{d_{max}} A_s \rho U_o^2 / \rho_s V_s$ where $C_{d_{max}}$ is the peak drag coefficient, and $A_s$ and $V_s$ are the sphere cross-sectional area and volume, respectively.

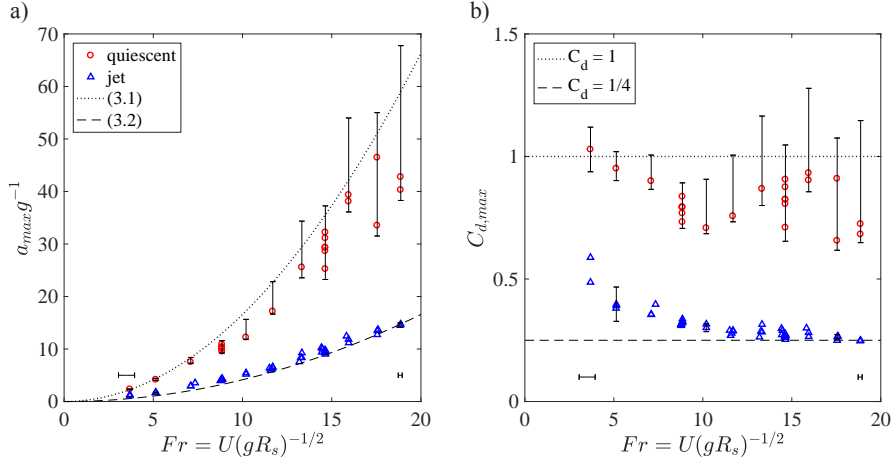

FIGURE 4. The maximum acceleration a) and drag coefficient b) of the initial impact peak are shown to be a function of $Fr$.

Non-dimensionalizing and rearranging the above equation we obtain

$$\frac{a_{max}}{g} = \frac{3}{8}\frac{\rho}{\rho_s}C_{d_{max}}Fr^2. \tag{3.1}$$

Based on the work of Shiffman & Spencer (1945) we take $C_{d_{max}} \approx 1$, which is reasonable for our spheres with density ratio $\rho_s/\rho = 2.26$ (see Fig. 4b). The dotted line in Fig. 4a plots (3.1) evaluated for the experimental conditions herein and is found to be a good approximation of the peak acceleration for spheres impacting on a quiescent surface. A large amount of uncertainty is found in the maximum impact acceleration for spheres impacting on quiescent pools at high velocities. This occurs because the duration of the impact peak becomes very short and the sampling frequency of the accelerometer is limited to 1000 Hz. Hence, the maximum acceleration during this peak occurs between data points leading to the large and asymmetric uncertainty bands for the quiescent impact cases with high velocity as illustrated in Fig. 4. Uncertainty is estimated using the $C_d$ verse depth curve from Shiffman & Spencer (1945) and considering the largest possible values our 1000 Hz sampling could have missed.

The spheres impacting inside a jet have lower peak accelerations for all $Fr$ compared to the quiescent impact cases (Fig. 4a). To rationalize these lower max accelerations, we note that the bottom of the jet cavity moves downwards with velocity $U_c$, which is about half the jet velocity $U_j$ (i.e., $U_c = \frac{1}{2}U_j$, and see Oguz et al. (1995); Speirs et al. (2018) for more details) during sphere impact, thus changing the relative impact velocity of the sphere ($U_{rel} = U_o - U_c$). For sufficiently large drop heights the distance that the jet falls before impact is approximately equal to the distance that the sphere falls before impact and thus, $U_o \approx U_j$ and $U_{rel} \approx \frac{1}{2}U_o$. Invoking the derivation of (3.1) using $U_{rel}$, the maximum acceleration of a sphere impacting behind a jet is

$$\frac{a_{max}}{g} \approx \frac{1}{4}\left(\frac{3}{8}\frac{\rho}{\rho_s}C_{d_{max}}Fr^2\right). \tag{3.2}$$

Therefore, if the sphere impacts a pool of water behind a jet the impact force can theoretically be reduced by up to 75% from the quiescent impact case. We can also examine the effect of the jet by looking at the change in the drag coefficient. Equation (3.2) can be rewrtitten as $a_{max}/g \approx \frac{3}{8}\frac{\rho}{\rho_s}C_{d_{max}}^{eqv}Fr^2$, where $C_{d_{max}}^{eqv} = \frac{1}{4}C_{d_{max}}$ is the equivalent drag coefficient. This implies that the maximum drag coefficient when impacting behind a jet is about a quarter that of a sphere impacting on a quiescent pool. Fig. 4b shows that the peak drag coefficient is significantly reduced when the sphere impacts in the wake of a jet.

Figure 3c shows that not only is the maximum impact acceleration lower for the jet cases, but the peak is much wider. This extended duration is quantified here by taking the peak width at half height $t_{hh}$, which is then made nondimensional using the sphere radius $R_s$ and the absolute impact velocity $U_o$ for the quiescent cases or the relative impact velocity $U_{rel}$ for the jet cases. When $t_{hh}U_o/2R_s$ or $t_{hh}U_{rel}/2R_s = 1$ this represents the time it takes for the sphere to be fully submerged in the water. For both cases this nondimensional time is found to be relatively constant for all impact velocities with $t_{hh}U_o/2R_s = 0.29 \pm 0.09$ for the quiescent cases and $t_{hh}U_{rel}/2R_s = 1.23 \pm 0.11$ for the jet cases. Hence, the peak in the sphere acceleration subsides before the sphere is fully submerged for the quiescent cases, but extends beyond the point of full submergence for the jet cases. We consider the effect of the extended peak duration on the total impulse of the sphere during this time. The magnitude of the total impulse is computed as $I = \int_0^{t_I} \rho_s V_s a_{total} dt$ by numerically integrating the acceleration curves in Fig. 3c; $t_I$ corresponds to the half height time following the maximum acceleration. Nondimensionalizing the impulse by the initial momentum $\rho_s V_s U_o$ for both cases reveals relatively constant values of $I/\rho_s V_s U_o = 0.067 \pm 0.003$ for the quiescent impact cases and $I/\rho_s V_s U_o = 0.175 \pm 0.006$ for the jet cases. Thus, although the spheres that impact inside a jet experience much smaller maximum acceleration magnitudes, the width of the initial impact peak is much larger leading to nearly three times the total impulse magnitude and, in turn, a larger reduction in velocity over the initial impact force duration.

To gain a better understanding of why the presence of a jet changes both the height and width of the initial acceleration peak, we use planar particle image velocimetry (PIV) to investigate the velocity field of the water under the sphere impact, a jet impact, and a sphere preceded by a jet (Fig. 5). The PIV images were taken with inter-frame spacing of 0.5 ms and processed with four passes at 64×64 pixel and two passes at 32x32 pixel interrogation regions using DaVis software. In Fig. 5a we see the first moments of impact, in which the sphere accelerates the fluid directly below itself ($t = 0.6$ ms). The velocity of the fluid directly in front of the sphere decreases from the sphere velocity to zero as the distance from the sphere increases. As the sphere descends further into the pool the mass of accelerated fluid in front of the sphere increases and the radius of the fluid mass stays approximately equal to the radius of the submerged portion of the sphere ($t = 1.2 - 2.4$ ms). When a jet of the same radius and velocity impacts, a larger local moving pool forms (Fig. 5b, note that the scaling of b differs from a). The velocity of the fluid directly in front of the jet cavity decreases from the jet velocity to zero as the distance from the cavity increases. This is the velocity field into which the sphere enters when impacting behind a jet, as shown in Fig. 5c. When the sphere first impacts the cavity bottom, the fluid that it passes through first has a velocity just smaller than its own and therefore the maximum impact acceleration is less than in the quiescent case. As the sphere continues to descend, the velocity of the fluid that it passes through gradually decreases. Hence the relative velocity between the sphere and the surrounding fluid gradually increases, extending the duration of impact.

Given that the primary source of the large initial impact force is added mass, and that the jet reduces this force by accelerating a mass of fluid in the pool, one would expect the amount of water contained in the jet to affect the peak accelerations. To attempt to approximate the required mass of fluid we start by asking the question: how much water in a jet is required to accelerate a large enough local moving pool with a velocity equal to $U_j/2$ (the cavity velocity)? Approximating the jet as a cylinder and maintaining the radius of the jet equal to $R_s$, we define the jet volume as $\pi R_s^2 L_{cr}$, where $L_{cr}$ is the critical or minimum jet length for maximum force reduction. We approximate the local moving pool as a hemisphere of radius $\kappa R_s$ and equate the momentum of the impacting jet with the momentum of the local moving pool as follows:

$$\rho U_j \pi R_s^2 L_{cr} \approx \rho \frac{U_j}{2} \frac{1}{2} \frac{4}{3} \pi (\kappa R_s)^3. \quad (3.3)$$

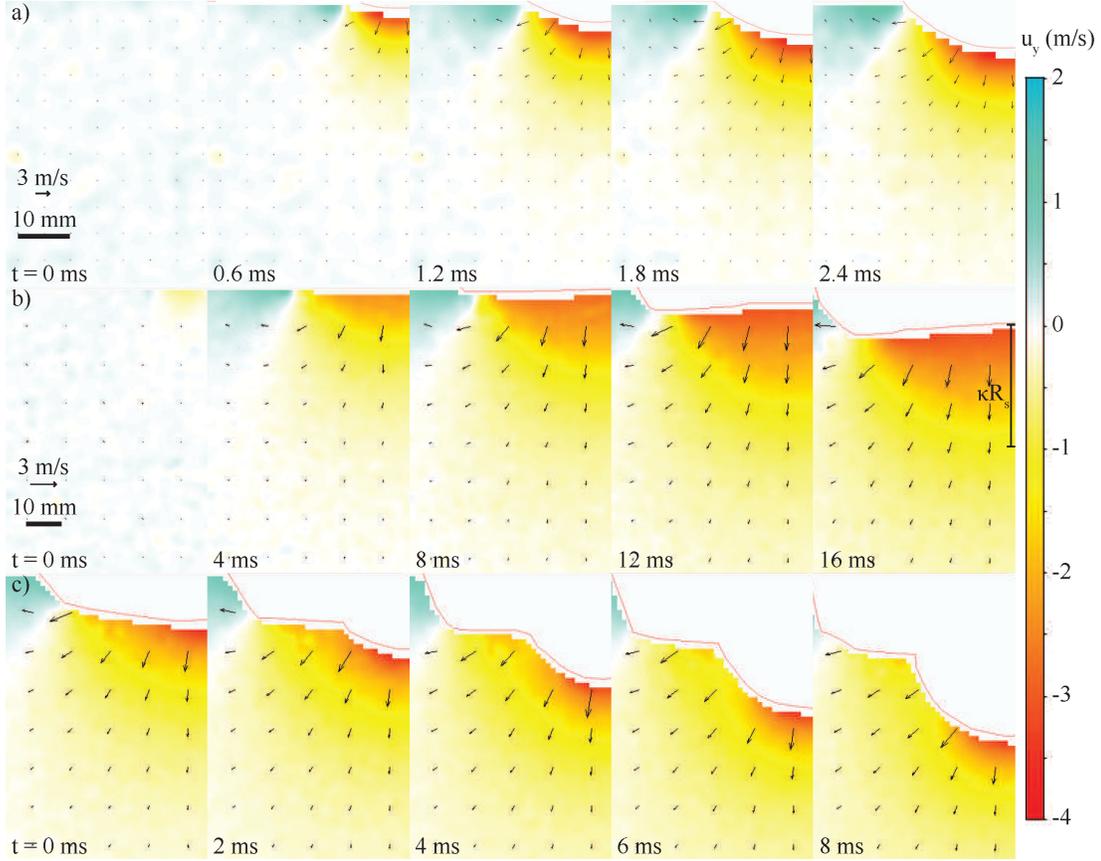

FIGURE 5. The flow fields created upon impact of a sphere, a jet and a jet followed by a sphere are shown in a) through c) respectively. The flow fields were measured using particle image velocimetry (PIV) and the thin red lines show the location of the masking, which covers the spheres and air cavities. The radii of the spheres and jets are 25 mm in each case. a) A sphere impacts an initially quiescent pool at 4.23 m/s accelerating the liquid in front of it. b) A jet with the same velocity impacts a quiescent pool, deforms and creates a large, local downward flow. c) A sphere at 4.45 m/s impacts the bottom of a cavity formed by a jet with the same impact conditions as in b). Only the left half of each image is shown. The length scale bar and velocity vector scale arrow shown in b) apply to c) as well. The coloring of the images shows the vertical velocity of the fluid $u_y$ with positive defined in the upward direction as shown in the color bar on the right. The bar in b) at $t = 16$ ms shows the radius of the local moving pool, $\kappa R_s$, used to predict $L_{cr}$. See supplemental movies 3 through 5.

Solving for $L_{cr}$ we find that $L_{cr} \approx \frac{1}{3}\kappa^3 R_s$. To approximate $\kappa$, we look at the velocity field in the pool created by the impact of a jet (Fig. 5b at $t = 16$ ms) and find the distance from the bottom of the cavity, along the axis of the jet, over which the average velocity equals $U_j/2$. Setting that distance equal to $\kappa R_s$ we find that $\kappa = 1.4$ which yields $L_{cr} \approx 0.91 R_s$ or approximately one sphere radius $R_s$.

To validate $L_{cr}$ experimentally we vary the length of the jet impacting in front of the sphere $L_j$ and examine its effect on the maximum sphere acceleration. To do this we plot the nondimensional jet length $L_j/R_s$ against the max acceleration experienced by the sphere impacting inside a jet $a_j$ normalized by the max acceleration experienced by a sphere impacting a quiescent pool $a_q$ at the same absolute velocity $U_o$. Fig. 6 shows that as $L_j$ increases from zero to about one sphere radius, the maximum impact acceleration decreases for all impact velocities tested ($U_o = 2.55, 4.23$, and $5.75$ m/s), but when $L_j \gtrsim R_s$ no further reduction is achieved. If $L_j = R_s$ the mass of liquid falling in front of the sphere is approximately equal to $V_s \rho / 2$, which is same as the added mass of a fully-submerged sphere. Thus, the most efficient jet that will reduce the force by 75% has mass on the order of the sphere's added mass, which makes sense when one considers that the added mass is the fluid accelerated by the sphere.

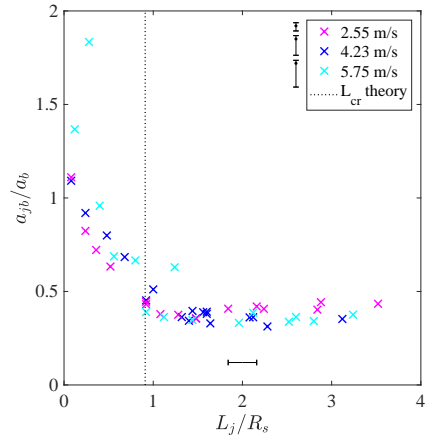

FIGURE 6. Increasing the jet length in front of an impacting sphere $L_j$ decreases the maximum acceleration experienced by the sphere during the very initial stages of impact for $L_j < R_s$. Once $L_j > R_s$ no further force reduction is achieved by increasing $L_j$. The maximum acceleration of a sphere impacting inside a jet $a_j$ is nondimensionalized by the max acceleration experienced by a sphere impacting a quiescent pool $a_q$ at the same absolute velocity $U_o$. The uncertainty of the nondimensionalized acceleration differs for each impact velocity, with the uncertainty bands shown next to the corresponding legend entry.

## 4. Conclusion

The water impact forces experienced by a falling body can be violent, due primarily to the fact that the body has to accelerate a mass of water from rest. If a liquid jet is made to impact prior to the body, then the forces can be significantly reduced by up to 75%. The jet accelerates the previously quiescent water thereby reducing the added mass effect on the impacting body. A jet length comparable to the sphere radius is sufficient to achieve this effect. This information could lead to a reduction in the impact force on objects that are dropped or launched into water such as torpedoes, sonobuoys, and space craft water landings.

### Acknowledgments

N.B.S., T.T.T. and J.B. acknowledge funding from the Naval Research Enterprise Internship Program (NREIP) and the Office of Naval Research, Navy Undersea Research Program (grant N0001414WX00811), monitored by Ms. Maria Medeiros. J.B. acknowledges funding from the Naval Undersea Warfare Center In-House Laboratory Independent Research program, monitored by Mr. Neil Dubois. N.B.S and T.T.T acknowledge funding from the Utah State University Research and Graduate Studies Development Grant Program.

REFERENCES

ARISTOFF, JEFFREY M. & BUSH, JOHN W. M. 2009 Water entry of small hydrophobic spheres. *Journal of Fluid Mechanics* **619**, 45–78.

BALDWIN, JOHN L 1971 Vertical water entry of cones. *Tech. Rep.*. Naval Ordnance Lab White Oak MD.

BODILY, KYLE G., CARLSON, STEPHEN J. & TRUSCOTT, TADD T. 2014 The water entry of slender axisymmetric bodies. *Physics of Fluids* **26** (7), 072108, arXiv: https://doi.org/10.1063/1.4890832.

COLEMAN, HUGH W & STEELE, W GLENN 2009 *Experimentation, validation, and uncertainty analysis for engineers*. John Wiley & Sons.

DUEZ, CYRIL, YBERT, CHRISTOPHE, CLANET, CHRISTOPHE & BOCQUET, LYDERIC 2007 Making a splash with water repellency. *Nat Phys* **3** (3), 180–183.

FALTINSEN, OM & ZHAO, RONG 1997 Water entry of ship sections and axisymmetric bodies. *Tech. Rep.* 827. High Speed Body Motion in Water.

GLASHEEN, J. W. & MCMAHON, T. A. 1996 Vertical water entry of disks at low froude numbers. *Physics of Fluids* **8** (8), 2078–2083, arXiv: https://doi.org/10.1063/1.869010.

GRADY, RJ 1979 Hydroballistics design handbook. *Naval Sea Systems command Hydromechanics Committee, January* .

GRUMSTRUP, TORBEN, KELLER, JOSEPH B. & BELMONTE, ANDREW 2007 Cavity ripples observed during the impact of solid objects into liquids. *Phys. Rev. Lett.* **99**, 114502.

KOROBKIN, AA & PUKHNACHOV, VV 1988 Initial stage of water impact. *Annual review of fluid mechanics* **20** (1), 159–185.

MAY, ALBERT 1975 Water entry and the cavity-running behavior of missiles. *Tech. Rep.*. Navsea Hydroballistics Advisory Committee Silver Spring MD.

MAY, ALBERT & WOODHULL, JEAN C. 1948 Drag coefficients of steel spheres entering water vertically. *Journal of Applied Physics* **19** (12), 1109–1121, arXiv: https://doi.org/10.1063/1.1715027.

MILOH, TOUVIA 1991 On the initial-stage slamming of a rigid sphere in a vertical water entry. *Applied Ocean Research* **13** (1), 43 – 48.

MOGHISI, M. & SQUIRE, P. T. 1981 An experimental investigation of the initial force of impact on a sphere striking a liquid surface. *Journal of Fluid Mechanics* **108**, 133–146.

OGUZ, HASAN N., PROSPERETTI, ANDREA & KOLAINI, ALI R. 1995 Air entrapment by a falling water mass. *Journal of Fluid Mechanics* **294**, 181–207.

SHEPARD, T, ABRAHAM, J, SCHWALBACH, D, KANE, S, SIGLIN, D & HARRINGTON, T 2014 Velocity and density effect on impact force during water entry of sphere. *J. Geophys. Remote Sens* **3** (129), 2169–0049.

SHIFFMAN, N & SPENCER, DC 1945 The force of impact on a sphere striking a water surface. *Tech. Rep.* No. AMG-NYU-133. New York Univ NY Courant Inst of Mathematical Sciences.

SPEIRS, NATHAN B., PAN, ZHAO, BELDEN, JESSE & TRUSCOTT, TADD T. 2018 The water entry of multi-droplet streams and jets. *Journal of Fluid Mechanics* **844**, 1084–1111.

THOMPSON, FL 1928 Water-pressure distribution on seaplane float. *Tech. Rep.* 290. National Advisory Committee for Aeronautics.

TRUSCOTT, TADD T., EPPS, BRENDEN P. & TECHET, ALEXANDRA H. 2012 Unsteady forces on spheres during free-surface water entry. *Journal of Fluid Mechanics* **704**, 173–210.

TVEITNES, T., FAIRLIE-CLARKE, A.C. & VARYANI, K. 2008 An experimental investigation into the constant velocity water entry of wedge-shaped sections. *Ocean Engineering* **35** (14), 1463 – 1478.

VON KARMAN, T 1929 The impact on seaplane floats during landing. *Tech. Rep.* 321. Natl. Advis. Comm. Aeronaut., Washington, DC.

WORTHINGTON, A. M. & COLE, R. S. 1900 Impact with a liquid surface studied by the aid of instantaneous photography. *Philosophical Transactions of the Royal Society of London. Series A, Containing Papers of a Mathematical or Physical Character* **194**, 175–199.

ZHAO, MENG-HUA, CHEN, XIAO-PENG & WANG, QING 2014 Wetting failure of hydrophilic surfaces promoted by surface roughness. *Scientific reports* **4**, 5376.